\documentclass[angeod,manuscript]{copernicus}  

\nolinenumbers
\usepackage{amsmath}
\usepackage{color}

\usepackage{natbib}

\begin{document}

\title{{Lorentzian Entropies and Olbert's $\kappa$ - distribution 
}}

\author[1,3]{R. A. Treumann}
\author[2]{W. Baumjohann}
\affil[1]{International Space Science Institute, Bern, Switzerland}
\affil[2]{Space Research Institute, Austrian Academy of Sciences, Graz, Austria}
\affil[3]{Geophysics Department, Ludwig-Maximilians-University Munich, Germany\\

\emph{Correspondence to}: Wolfgang.Baumjohann@oeaw.ac.at
}

\runningtitle{Olbert distribution and entropy}

\runningauthor{R. A. Treumann \& W. Baumjohann}

\received{ }
\pubdiscuss{ } 
\revised{ }
\accepted{ }
\published{ }


\firstpage{1}

\maketitle

\begin{abstract}
This note derives the various forms of entropy of systems subject to {Olbert distributions (generalized Lorentzian probability distributions known as $\kappa$-distributions) }which are  frequently observed particularly in high temperature plasmas. The general expression of the partition function in such systems is given as well in a form similar to the Boltzmann-Gibbs probability distribution, including a possible exponential high energy truncation. We find the representation of the mean energy as function of probability, and provide the implicit form of {Olbert (Lorentzian) entropy} as well as its high temperature limit. The relation to phase space density of states is obtained. We then find the entropy as function of probability, an expression which is fundamental to statistical mechanics and  here to its Olbertian version. Lorentzian systems through internal collective interactions cause correlations which add to the entropy. Fermi systems {do not obey Olbert} statistics, while Bose systems might at temperatures sufficiently far from zero.
    

\end{abstract}
\section{Introduction}
Many-particle systems not in equilibrium like high-temperature plasmas are usually subject to kinetic theory \citep[cf., e.g.,][]{klimontovich1967}. In equilibrium or stationary quasi-equilibrium, obeying a very large number of degrees of freedom, they can beneficially be treated by the probabilistic methods of statistical mechanics. Conventional textbook knowledge \citep{kittel1980} tells that, for the micro-canonical system under consideration being in thermal exchange with a large thermal bath at temperature $T\equiv\beta^{-1}$ (here taken in energy units), the probability $p_\alpha$ of finding it in some particular energy state $E_\alpha$ is proportional to the Boltzmann factor $p_\alpha\propto\exp (-\beta E_\alpha)$. The sum of all un-normalized probabilities of the $\alpha$ states is the partition function $Z=\sum_\alpha p_\alpha=\sum_\alpha\exp(-\beta E_\alpha)$, and the normalized Gibbs probability for the state $\alpha$ becomes
\begin{equation}\label{eq-gibbs}
P_\alpha = Z^{-1}\exp (-\beta E_\alpha)
\end{equation}
The partition function $Z\equiv Z(\beta,\{V\})$ is a function of $\beta$ and all constraining parameters $\{V\}$ which determine the state $\alpha$, a property that enables calculating a number of thermodynamically interesting average quantities of the system. Varying the constraints $\{V\}$ implies that work is done on the system. 

Observations in space plasma physics \citep[for examples cf.,][]{christon1988,christon1989,christon1991} as well as in other high temperature systems indicate that the probability distribution of particles (charged or neutral)  in a set of energy states $E_\alpha$ deviates from the classical bell (respectively gaussian) shape, frequently exhibiting quasi-stationary power law tails $P_\alpha\propto E_\alpha^{-\kappa}$ for $E_\alpha> \beta^{-1}$, possibly cut off exponentially at large energy. Probability distributions of this kind of family, known as $\kappa$-distributions \citep[introduced\footnote{Stanislaw (Stan) Olbert (1923-2017, of Polish origin, after WW II a graduate of Arnold Sommerfeld in Munich, and since 1957 Professor of Physics  at MIT, working on the American Space Program with Bruno Rossi, the main discoverer of the X-ray sky and, together with Riccardo Giacconi, who later was awarded the Nobel Prize for this, founder of X-ray astronomy) invented the $\kappa$-probability distribution to fit observed IMP spacecraft particle spectra. He suggested its application to electron fluxes measured by the OGO spacecraft to \citet{vasyliunas1968} whose publication became one of the most referenced papers in space physics, for no other reason than the first refereed formal appearance of {Olbert's} $\kappa$ distribution in the literature.}  by][]{olbert1967}, have been widely discussed in the literature \citep[for a review see, e.g.,][and references therein]{livadiotis2013}. In the following we refer to them as \emph{Olbert's $\kappa$-distributions} or simply {\emph{Olbert's distribution}}.  General physical arguments for their existence as stationary states far from thermal equilibrium were given \citep[first by][]{treumann1999,treumann2008}. Direct weak turbulence calculations of plasma-electron momentum distributions  by \citet[][in interaction with a photon bath]{hasegawa1985} and by \citet[][{accounting for spontaneous and induced emission as well as absorption of Langmuir waves}]{yoon2005,yoon2006} partially reproduced $\kappa$-distributions in the long term limit,  suggesting that under quasi-stationary conditions nonlinear equilibria can be produced with $\kappa$-distributions being their probabilistic signature. 

\section{Lorentzian Generalization}
In generalizing the classical statistical mechanics we start from a Lorentzian modification of the Boltzmann factor which leads to the {Olbert probability distribution} known as $\kappa$-distribution.

\subsection{Boltzmann-Olbert distribution}
In fact, the Boltzmann factor, being at the heart of Gibbs' normalized probability, is the large $\kappa$ limit of a more general Lorentzian, the Olbert $\kappa$-probability function 
\begin{equation}\label{eq-kappadist}
P_{\kappa\alpha}(E_\alpha,\beta)=Z^{-1}_{\kappa, r}\Big[1+\frac{\beta E_\alpha}{\kappa}\Big]^{-(\kappa+r)}, \qquad \lim_{\kappa\to\infty}P_{\kappa\alpha}\rightarrow P_\alpha
\end{equation}
(with $r=$ const $\neq0$) as can easily be confirmed applying l'Hospital's rule. It corresponds to the above mentioned, experimentally  frequently confirmed $\kappa$-distribution. The resulting Olbert-partition function $Z_\kappa$ is, in analogy to Gibbs' partition function, defined as
\begin{equation}
Z_{\kappa,r}(\beta)=\sum_\alpha\Big[1+\frac{\beta E_\alpha}{\kappa}\Big]^{-(\kappa+r)}
\end{equation}
It warrants that the Olbert probabilities of states $\alpha$ are normalized and add up to $\sum_\alpha P_{\kappa\alpha}(E_\alpha,\beta)=1$. Performing this sum requires knowledge of the different energy states $E_\alpha$, which in general  cannot be done easily. In the following we show that assuming this form, the rules of classical statistical mechanics can be made applicable to the Olbert-Lorentzian with only weak modifications.

\subsection{{Remark on convergence}}
Before proceeding, we {briefly refer to the convergence of Olbert's} $\kappa$-probability distribution Eq. (\ref{eq-kappadist}). 

{The Olbert probability converges for arbitrary power $\kappa>0$. It does, however, for constant $\kappa$ not allow the calculation of arbitrarily high average moments, for instance if one is interested in fluid descriptions. [In principle, at this stage $\kappa(\beta, E_\alpha)$ being a function of temperature and/or even energy states $E_\alpha$, is not excluded; in the latter case one would, however, require that its dependence is weak, in order to maintain the above summation procedure as simple as possible. Such a dependence is implicit to the  nonlinear calculations of \cite{hasegawa1985} and \cite{yoon2005}.] The additional freedom introduced by the constant $r$ just adjusts for the mean energy in an ideal gas \citep[see, e.g.,][and references therein]{treumann2014}. In general, however, the number of moments which can be calculated is limited. In a fluid approach, it requires artificially truncating the chain of moments, for instance by applying a water-bag model for $\kappa(\beta, E_\alpha)$ of the kind $\kappa=\mathrm{const}, E_\alpha\leq E_c$,  and $\kappa\to\infty, E_\alpha>E_c$ \citep[implicitly assumed in][]{treumann2004}. Truncation may be justified via additional assumptions on the underlying physics, like suppression of higher moments than heat flows and similar conditions. Physically this may not be unreasonable. From a formal point of view, brute force truncation is not satisfactory.} However, this restriction can easily be circumvented \citep[see e.g.,][]{scherer2017,lazar2020,treumann2004} when introducing an exponential cut-off energy $\beta E_c\gg 1$ through
\begin{equation}\label{eq-kappadist-a}
P_{\kappa\alpha}=\frac{e^{-E_\alpha / E_c}}{Z_{\kappa, r}}\Big[1+\frac{\beta E_\alpha}{\kappa} \Big]^{-(\kappa+r)}, \qquad \beta E_c\gg 1
\end{equation} 
which warrants convergence of all moments for arbitrary $\kappa>0$ \citep{scherer2017,lazar2020}. {The chain of physically interesting moments is discussed in these papers. In the Olbert partition function the energy cut-off simply appears as a truncator}
\begin{equation}
Z_{\kappa,r}=\sum_\alpha e^{-{E_\alpha/E_c}}\Big[1+\frac{\beta E_\alpha}{\kappa}\Big]^{-(\kappa+r)}
\end{equation}
{not having any further effect on the determination of averages and/or any other thermodynamic quantities than warranting the convergence of the chain of moments. The independence of the exponential cut-off on temperature and $\beta$ guaranties that in all derivatives or integrals with respect to $\beta$ it appears as an energy dependent factor. An example has been given \citep{treumann2018} by application to the Cosmic Ray energy spectrum, where the cut-off is, for quantum physical reasons, found in the GZ-energy spectral limit. Its inclusion, if necessary, does not cause any principal problems. In the following we therefore suppress it in order to not unnecessarily complicate the expressions.}

\section{Olbert-Lorentzian statistics}
It is reasonable to assume that, given the above definition of the probability, ensemble averages can be calculated as linear mean values, with the probability $P_{\kappa\alpha}$ determining the weight, each energy level contributes. This is the basic probability assumption. One may argue that this may not necessarily be true if the probabilites of the states are not independent. Such arguments have been put forward in some entropy definitions \citep[see, e.g.,][]{wehrl1978} some of them like Renyi and Tsallis $q$-entropies   \citep[cf.][for their invention]{balatoni1956,renyi1970,tsallis1988} are used in chaotic theory.\footnote{Historically,  Renyi's proposal of a $q$-entropy \citep[for the complete theory see][]{renyi1970}  came first \citep[in a badly accessible publication by][in very general form, which implicitly already contained Tsallis' entropy as a particular case]{balatoni1956}. This might have been known to Stan Olbert (who probably was familiar with the Hungarian literature). He used the property that for large parameter $q\to\infty$, as proposed by Renyi, the modified mathematical expression agreed with Boltzmann's exponential. Olbert, however, tried a substantially simpler analytical form, calling the free parameter $\kappa$ instead of $q$ to distinguish it from Renyi's logarithm, as in fact it has different meaning. Two decades later, \citet{tsallis1988}  used the property of Olbert's function, presumably not knowing  Olbert's or Vasyliunas' so much earlier papers, and probably also not that by Balatoni and Renyi, though Renyi  referred to the latter in his book, which Tsallis should have been familiar with, because, at that time, Renyi's $q$-entropy was already highly celebrated in the then blossoming chaos theory. In contrast to Olbert however, Tsallis did not apply it to the probability distribution. Rather, following Renyi's logarithmic approach, he used it in the entropy definition, arriving at his analytically simpler modified $q$-entropy. The two approaches of Olbert and Tsallis thus differ in the way of how the substitution for the exponential is used. As ours, Olbert's interest was in the observed probability or momentum space distribution and thus manifestly practical. Renyi's interest and later Tsallis' were theoretical and thus in the entropy. Tsallis' led, consequently developed, to  his thermostatistics. In contrast, in an atempt to justify Olbert's distribution, we arrived originally at a $\kappa$-distribution from a consequent reference to kinetic theory \citep{treumann1999}, not yet recognizing, however, the important role of the constant $r$. As it turns out, both approaches are indeed rather different, even though a formal relation between the parameters $\kappa$ and $q$ can easily be construed while maintaining their different meanings, which is frequently overlooked when identifying $q$ and $\kappa$ statistics, as these have little in common.} However, as long as there is no need to worry about, the $\kappa$-generalization of the probabilities already accounts for a particular kind of internal correlations among the occupations of the different states. The states are physically ordered as in Boltzmann-Gibbs theory, while the probabilities of their occupations have become not completely independent. In this spirit the mean energy is defined as
\begin{equation}
U(\beta)\equiv\big\langle E\big\rangle =Z^{-1}_{\kappa,r}\sum_\alpha E_\alpha\Big[1+\frac{\beta E_\alpha}{\kappa}\Big]^{-\kappa-r}
\end{equation}

\subsection{Mean Energy}
In full generality the sum can formally be done in two ways when observing the properties of the partition function. The first way completes the energy, and one easily finds that
\begin{equation}\label{eq-u1}
U(\beta) = \frac{\kappa}{\beta}\Big[\frac{Z_{\kappa,r-1}(\beta)}{Z_{\kappa,r}(\beta)}-1\Big]
\end{equation}
also showing the importance of having made use of the freedom of introducing the arbitrary constant $r\neq0$. 

A second form, resembling that of conventional statistical mechanics, takes advantage of the differential property 
\begin{equation}
\frac{\partial Z_{\kappa,r-1}}{\partial\beta}=-\frac{\kappa+r-1}{\kappa} Z_{\kappa,r}U(\beta)
\end{equation}
of the partition function, yielding 
\begin{equation}\label{eq-u2}
U(\beta)= -\frac{\kappa}{\kappa+r-1}\frac{Z_{\kappa,r-1}}{Z_{\kappa, r}}\Big(\frac{\partial\log Z_{\kappa,r-1}}{\partial\beta}\Big)_{\{V\}}
\end{equation}
Here, generalization to Olbert-Lorentzian distributions introduces the (inconvenient) partition function ratio of different indices. It again shows the need for the additional constant $r\neq 1$ 
which depends on the assumptions on an underlying model. For instance under classical ideal gas conditions, with continuously distributed energy states, the average thermal energy (in three dimensions and isotropy) is $\beta U=\frac{3}{2}$.  On switching to momentum  $\mathbf{p}$ with $E_\alpha\to p^2/2m$ and integrating over momentum space, one obtains   that $r=\frac{5}{2}$ in this particular case \citep[][and elsewhere; see references therein]{treumann2014}. This is not necessarily true, however, for discrete energy levels $E_\alpha$ in more general non-ideal or quantum conditions. There $r$ must be chosen differently and a general prescription for its choice cannot be given \emph{a priori}.

Both the above forms  apply to any micro-canonical $\kappa$-system. Generalization to canonical systems is easily done in the same way as in statistical mechanics \citep[cf., e.g.,][]{kittel1980} via introducing the dependence on (possibly variable) particle number $N$. It requires reference to Lagrange multipliers  $\mu$ playing the role of  chemical potentials for each subsystem, and transforming $E_\alpha\to E_\alpha-\mu$ in the probabilities and partition functions. Clearly,  all $\mu$ must become equal in thermal equilibrium.

The two Eqs. (\ref{eq-u1}, \ref{eq-u2}) allow for the determination of the ratio of the partition functions by eliminating $U(\beta)$
\begin{equation}
\frac{Z_{\kappa,r-1}}{Z_{\kappa, r}}=\frac{\kappa+r-1}{\kappa+r-1+\beta\Big(\partial\log Z_{\kappa, r-1}/\partial\beta\Big)_{\{V\}}}
\end{equation}
This is a recursive relation between the $\kappa$ partition functions. Combined with Eq. (\ref{eq-u2}), it gives a final expression for the mean energy
\begin{equation}\label{eq-ufin}
U(\beta)=-\frac{\kappa\Big(\partial\log Z_{\kappa, r-1}/\partial\beta\Big)_{\{V\}}}{\kappa+r-1+\beta\Big(\partial\log Z_{\kappa, r-1}/\partial\beta\Big)_{\{V\}}}
\end{equation}
which contains just the $r$-reduced partition function. Like in ordinary statistical mechanics, $U(\beta)$ is determined as a derivative form of the partition function. {This shows that all other statistical mechanical quantities can be derived solely from the partition function which therefore contains all the physics of the micro-canonical system.} Still being quite involved, this form, as expected for very large $\kappa$, coincides with the expression $U= -\big[\partial(\log Z)/\partial\beta\big]_{\{V\}}$ of the mean energy in Boltzmann-Gibbs statistical mechanics. It is thus consistent with the expectations. Moreover, at  large temperatures $\beta\to0$, and the mean classical energy  becomes
\begin{equation}\label{eq-uhight}
U(\beta)=-\frac{\kappa}{\kappa+r-1}\Big[\frac{\partial(\log Z_{\kappa,r-1})}{\partial\beta}\Big]_{\{V\}}\qquad T\gg0
\end{equation}

The general second last equation (\ref{eq-ufin}), which holds for arbitrary $\beta<\infty$, resolved for the derivative of the partition function as function of mean energy $U(\beta)$, yields
\begin{eqnarray}\label{eq-logzu}
\Big(\frac{\partial \log Z_{\kappa,r-1}}{\partial\beta}\Big)_{\{V\}}&=&\Big(1+\frac{r-1}{\kappa}\Big)U(\beta)\Big(1+\frac{\beta U(\beta)}{\kappa}\Big)^{-1}\\
&=&\Big(1+\frac{r-1}{\kappa}\Big)\Big[1-\Big(1+\frac{\beta U(\beta)}{\kappa}\Big)^{-1}\Big]
\end{eqnarray}
an expression which can be made use of later. At high temperatures, i.e. small $\beta$ the first version shows that the derivative of the partition function yields the mean energy, the usual Boltzmann-Gibbs result. 

At very low temperature $\beta\gg1$ the mean energy drops out, which contradicts the physical intuition showing that the theory in this form applies only to temperatures far from zero.  The logarithm of the partition function is the integral
\begin{equation}
\log Z_{\kappa,r-1}=\Big(1+\frac{r-1}{\kappa}\Big)\Big[\beta-\int \frac{d\beta}{1+\beta U(\beta)/\kappa}\Big] +G_\kappa(\{V\})
\end{equation}
with $G_\kappa(\{V\})$ a function of the constraints alone. Olbert-Lorentzian statistical mechanics in the above form applies  to micro-canonical systems at high temperature only. It does, in this form not describe quantum systems consisting of many components, a conclusion we had drawn already earlier from different reasoning. 
This conclusion may however be circumvented when large external potential fields $\Phi$ are imposed, for instance strong electric \citep[cf., e.g.,][who try an application to high temperature no-ideal quantum systems ]{domenech-garret2015} or gravitational potential fields (an example would be the region around the black hole horizon) in which case the difference $U(\beta)-\Phi>0$ may become positive for $-\Phi>\kappa/2\beta$.

\subsection{Entropy}
The most important quantity in statistical mechanics is the entropy. Differentiating the energy $U(\beta)$ with respect to temperature $k_B\beta^{-1}$ while fixing the set of constraints $\{V\}$ gives the heat capacity 
\begin{equation}
C_{\{V\}\kappa} = -k_B\beta^2\Big(\frac{\partial U(\beta)}{\partial\beta}\Big)_{\{V\}}
\end{equation}
With entropy $S$, one has quite generally $TdS = C_{\{V\}}dT$ holding in the micro-canonical ensembles where the volume is fixed, $dV=0$. Hence $C_{\{V\}}=-\beta(\partial S/\partial\beta)_{\{V\}}$. Keeping the constraints fixed, these relations  as usually lead to
\begin{equation}
\Big(\frac{\partial S}{\partial\beta}\Big)_{\{V\}}=-k_B\beta\Big(\frac{\partial U(\beta)}{\partial\beta}\Big)_{\{V\}}
\end{equation}
Integration with respect to $\beta$ then yields in full generality the well known formal expression for the wanted Olbert entropy $S_\kappa$ of the micro-canonical system 
\begin{equation}\label{eq-entropy}
\frac{S}{k_B}=-\beta U(\beta)+\int d\beta \, U(\beta) + G_S(\{V\})
\end{equation}
as the integral over the mean energy $U(\beta)$, where $G_S(\{V\})$ is an arbitrary function of the constraints alone. In classical statistics this formula yields the well-known closed analytical expression of the entropy.  Unfortunately, {in Olbert's case} the mean energy Eq. (\ref{eq-ufin}) is not as simple as in Boltzmann-Gibbs statistical mechanics.\footnote{At this point $\kappa$ and $q$ statistics differ for the simple fact that in the latter the entropy is analytically prescribed in the form of a rational function with real $q$, and the complication is transferred to the construction of the distribution. {Here, instead, the starting point is the observed distribution which, naturally, leads to complications in finding the entropy, as it is the entropy which contains the complicated physics, which then leads to the measured probability distribution. One should also note that the combined entropies of two systems in both cases, $q$ and Olbert statistical mechanics, though the two theories are different and describe different physics, are super-additive, sometimes called non-extensive. They contain an additional mixed term which contributes to the entropy, as criticized by \citet{nauenberg2003}. This, however, does not mean that the theory has no physical meaning. It just implies that the theory describes statistical quasi-equilibria far from thermal equilibrium, i.e. slowly variable quasi-stationary states which pass through several equilibria, typical for non-equilibrium statistical mechanics \citep{landau1980}.}} We are thus stuck for the moment. Nevertheless, taking the derivative of the mean energy with respect to $\beta$, one obtains formally
\begin{equation}\label{eq-entropyderivativ}
\Big(\frac{\partial S}{\partial\beta}\Big)_{\{V\}}=-\kappa k_B\beta\frac{\partial}{\partial\beta}\Big[\frac{\partial(\log Z_{\kappa,r-1})/\partial\beta}{\kappa+r-1+\beta\partial(\log Z_{\kappa,r-1})/\partial\beta}\Big]_{\{V\}}
\end{equation}
as an implicit  expression for the derivative of the entropy $S$ as functional of the partition function. It replaces the corresponding relation in classical Boltzmann-Gibbs statistical mechanics which applies to any purely stochastic many particle system, in particular to high-temperature plasmas.
 
Eq. (\ref{eq-entropy}) is the entropy of a micro-canonical $\kappa$ system. It is a quite involved form whose properties cannot be easily inferred. Its discussion requires the complete knowledge of the set of energy levels of the micro-canonical system. As discussed above, its extension to canonical systems is straightforward, as well as the inclusion of an exponential ``ultraviolet'' truncation of the distribution at high energy $E_c>U$. All interesting statistical mechanical properties of the $\kappa$ ensemble can \emph{in principle} be deduced from this entropy respectively the partition function $Z_{\kappa,r-1}$.

\subsection{High-temperature limit}
In the high temperature small $\beta$ limit one neglects the derivative in the denominator in the second last equation. In this case the entropy becomes a $\kappa$-modified (Boltzmann-Olbert-Lorentzian) entropy
\begin{equation}
\Big(\frac{\partial S}{\partial\beta}\Big)_{\{V\}}=-\frac{\kappa k_B\beta}{\kappa+r-1}\frac{\partial}{\partial\beta}\Big[\frac{\partial(\log Z_{\kappa,r-1})}{\partial\beta}\Big]_{\{V\}},\qquad T\gg0
\end{equation}
No zero-temperature expression exists, while the role of the partition function is played by the sum of the probabilities of the states indexed by the constant power $r-1$ instead of $r$. For a three-dimensional ideal gas with continuous energy spectrum one has $r=\frac{5}{2}$, and its high-temperature classical $\kappa$-partition function is 
\begin{equation}
Z_{\kappa+\frac{3}{2}}(\beta)=\sum_\alpha p_{\alpha,\kappa+\frac{3}{2}}\equiv\sum_\alpha\Big(1+\frac{\beta E_\alpha}{\kappa}\Big)^{-\kappa-\frac{3}{2}}, \qquad T\gg 0
\end{equation}
With this partition function and the definition of the high-temperature mean energy (\ref{eq-uhight}) we are in the position to obtain the high-temperature entropy in the form in which it applies to fluids and plasmas:
\begin{equation}
\frac{S_\kappa}{k_{B\kappa}}=-\beta\Big[\frac{\partial\log Z_{\kappa+r-1}(\beta)}{\partial\beta}\Big]_{\{V\}}+\log Z_{\kappa+r-1}(\beta) 
\end{equation}
where we left $r$ undetermined available for application to any non-ideal systems, and dropped the arbitrary function $G_S$ of the constraints which can be added when needed, for instance to account for boundary conditions. Except for the modification of the partition function, the entropy at high temperatures is measured in units of a $\kappa$-reduced Boltzmann constant 
\begin{equation}
k_{B\kappa}=k_B\Big(1+\frac{r-1}{\kappa}\Big)^{-1},\qquad 0\nleq\kappa<\infty
\end{equation}
which in a three-dimensional ideal plasma becomes $k_{B\kappa}=k_B/(1+3/2\kappa)$. 

\subsection{Phase space density of states}
As in ordinary Boltzmann-Gibbs statistical mechanics, the Olbert partition function for large numbers of states (energy levels) $\Omega_\kappa$, which is the volume of the phase space, is well approximated by
\begin{equation}
Z_{\kappa, r-1}\approx \Omega_{\kappa,r-1}\Big(1+\frac{\beta U(\beta)}{\kappa}\Big)^{-\kappa-r+1}
\end{equation}
the product of the phase space volume $\Omega_{\kappa,r-1}$ and the probability of the most probable state, which is the state of mean energy $U(\beta)$. This holds, in particular for the exponentially truncated distribution, because the high energy states contribute very little if only the energy fluctuations are not overwhelmingly large. These fluctuations become large only in systems containing very small numbers of particles, which is barely given at the assumed high temperatures in a plasma. 

Taking advantage of the dependence of the ratio of the partition functions $Z_{\kappa,r-1}/Z_{\kappa, r}$ on the average energy $U(\beta)$, which does not depend on $r$, one  finds that
\begin{equation}
\Omega_{\kappa,r-1}=\Omega_{\kappa,r}\equiv\Omega_\kappa
\end{equation}
At high temperatures $\beta\ll 1$, we have
\begin{equation}
S_{\kappa}\approx k_{B\kappa} \log Z_{\kappa,r-1}+k_{B\kappa}(\kappa+r-1)\log\Big(1+\frac{\beta U(\beta)}{\kappa}\Big)\approx k_{B\kappa}\log\Omega_{\kappa}
\end{equation}
which can also be written
\begin{equation}
P_{\kappa}\sim\Omega_{\kappa}=\exp\Big(S_{\kappa}/k_{B\kappa}\Big)
\end{equation}
In classical high-temperature micro-canonical systems (many-particle plasmas) this closes the circle, as we have shown \citep{treumann2008} that from this equation it follows by standard methods that the probability distribution is given by Eq. (\ref{eq-kappadist}). Generalization to the canonical system of $N$ particles is straightforward. Notably,  it generalizes  to $\kappa$-systems Einstein's prescription \citep{einstein1905} of the dependence of the phase space density on entropy in his proof of the stochastic nature of the diffusion in Brownian motion, however with substantially more complicated expression for the entropy. 

\subsection{Approximation}

The Olbert $\kappa$-distribution maintains the structure of statistical mechanics at high temperatures while it substantially modifies it at moderate and low temperatures, with no zero temperature limit existing. We have argued previously that this is quite reasonable whenever internal correlations come into play causing $\kappa$ to deviate strongly from $\kappa=\infty$. Classically this can  happen only at large $T$  and is due to nonlinear interactions which violate ideal stochasticity and cause anomalous effects like anomalous diffusivity.  Nevertheless, in the range 
\begin{equation}
\log Z_{\kappa,r-1}\gg (\kappa+r-1)\log\beta \quad\mathrm{or}\quad Z_{\kappa,r-1} \gg \beta^{\kappa+r-1}
\end{equation}
which holds for sufficiently large $\beta$, the equation for the Olbert entropy simplifies. In this case the derivatives of the logarithms of the partition function cancel, and the entropy equation becomes
\begin{equation}
\Big(\frac{\partial S}{\partial\beta}\Big)_{\{V\}} \approx -\kappa k_B\beta\frac{\partial}{\partial\beta}\frac{1}{\beta},\qquad 1\ll \beta<\infty
\end{equation}
which of course holds for finite $\beta$ only. Integration then yields that in this $\beta$ range 
\begin{equation}
\frac{S}{k_B}\propto \kappa\log\beta \ll \frac{\kappa}{\kappa+r-1}\log Z_{\kappa,r-1}
\end{equation}
or otherwise
\begin{equation}
Z_{\kappa,r-1}\gg \exp\frac{S}{k_{B\kappa}}
\end{equation}

In the moderate temperature range where $0\ll \beta\ll\infty$ no closed forms for either the energy nor the entropy are obtained. For those values of $\beta$ the full expression (\ref{eq-entropyderivativ}) for the derivative of the entropy applies. A correction to this equation follows when taking the first next term of the expansion of the denominator
\begin{equation}
\Big(\frac{\partial S}{\partial\beta}\Big)_{\{V\}}=-\frac{\kappa k_B}{\kappa+r-1}\beta\frac{\partial}{\partial\beta}\Big\{\Big(\frac{\partial\log Z_{\kappa,r-1}}{\partial\beta}\Big)_{\{V\}}-\frac{\beta(\partial\log Z_{\kappa,r-1}/\partial\beta)^2
_{\{V\}}}{\kappa+r-1}+\mathrm{h.o.t.}\Big\}
\end{equation} 
The first term yields, when integrated, the above high-temperature entropy. The second term is quadratic and hence remains to be negative. It subtracts from the first term. Reducing the temperature, i.e. increasing $\beta$  obviously diminishes the derivative of the entropy, because the last quadratic term is always positive. It seems that the derivative of the entropy as function of temperature in a $\kappa$-system flattens out when the temperature drops into the intermediate range. Any  $\kappa\neq\infty$ affects the increase in entropy. 

{In principle the last equation can be solved iteratively for the entropy which then retains the effects of the parameter $\kappa$ outside the ranges of very large and small $\beta$.}

\section{Entropy as functional of probability}
Boltzmann defined the micro-canonical entropy $S_{B\alpha}\propto \log p_\alpha$ as a functional of probability. {The average measured entropy is its expectation value, the sum $\langle S_B\rangle\propto \sum_\alpha p_\alpha\log p_\alpha$ of all probability-weighted contributions of the states to the entropy. For  a continuous distribution of states this is as usually defined as the probability integral  taken over the micro-canonical entropy. }For arbitrary temperatures the above expression cannot be integrated to provide a general analytical form for the entropy comparable to conventional Boltzmann-Gibbs statistical mechanics. This was possible only at high temperatures. One can, however attempt to find an expression for the functional dependence of the entropy on the probability in order to have an equivalent representation to the Boltzmann-Gibbs entropy {when dealing with the Olbert entropy.} 

\subsection{Reformulation of mean energy and entropy}
For this to achieve the mean energy (\ref{eq-u1}) must be rewritten in terms of the probability $p_{\kappa\alpha}(\beta)$. {This can, indeed, be done, and} the corresponding expression reads
\begin{equation}\label{eq-inverse}
U\{p_{\kappa\alpha}(\beta)\}=\frac{\kappa}{\beta}Z^{-1}_\kappa\{p_{\kappa\alpha}(\beta)\}\sum_\alpha \Big[p_{\kappa\alpha}^{1-\gamma}(\beta)-p_{\kappa\alpha}\Big]
\end{equation}
where we introduced the exponent
\begin{equation}
\gamma\equiv \frac{1}{\kappa+r}
\end{equation}
The braces indicate the functional dependence on $p_{\kappa\alpha}$. In fact Eq. (\ref{eq-inverse}) is identical to the inverse function which has been proposed \citep{treumann1999,treumann2014} in the particular case of the {Olbert-Lorentzian probability} distribution.\footnote{In other approaches this inverse appears as a mysterious ``escort distribution'' which plays the role of some integration condition when forming lowest order moments. In fact it is nothing but an inverse function as was proposed already \citep{treumann2014} and in other choices of the probability distribution would be obtained in the same way by inversion.}  On use of this functional dependence in the expression for the  $\beta$-derivative we have 
\begin{equation}
\frac{1}{\kappa k_B}\Big(\frac{\partial S}{\partial\beta}\Big)_{\{V\}}=-\sum_\alpha\beta\Big\{\frac{\partial}{\partial\beta}\frac{1}{\beta} Z^{-1}_{\kappa}\{p_{\kappa\alpha}(\beta)\}\Big[p_{\kappa\alpha}^{1-\gamma}(\beta)-p_{\kappa\alpha}(\beta)\Big]\Big\}_{\{V\}}
\end{equation}
which shows that the derivative of the micro-canonical entropy with respect to $\beta$ respectively temperature $T$ is the sum over all states of the particular entropies 
\begin{equation}
\frac{1}{\kappa k_B}\Big(\frac{\partial S_\alpha}{\partial\beta}\Big)_{\{V\}}=-\beta\Big\{\frac{\partial}{\partial\beta}\frac{\big[p_{\kappa\alpha}^{1-\gamma}(\beta)-p_{\kappa\alpha}(\beta)\big]}{\beta Z_{\kappa}\{p_{\kappa\alpha}(\beta)\}}\Big\}_{\{V\}}\end{equation}
of the $\alpha$ states. This expression  replaces Boltzmann's definition to become Olbert's micro-canonical entropy, and it follows that
\begin{equation}\label{eq-entr}
\frac{S_\alpha\{p_{\kappa\alpha}(\beta)\}}{\kappa k_B}=-\frac{p_{\kappa\alpha}(\beta)}{Z_{\kappa}\{p_{\kappa\alpha}(\beta)\}}\big[ p_{\kappa\alpha}^{-\gamma}(\beta)-1\big] + \int d\beta\frac{p_{\kappa\alpha}(\beta)}{Z_{\kappa}\{p_{\kappa\alpha}(\beta)\}}\frac{\big[ p_{\kappa\alpha}^{-\gamma}(\beta)-1\big] }{\beta}
\end{equation}
The factor $p_{\kappa\alpha}/Z_\kappa=P_{\kappa\alpha}$ is the normalized Olbert-Gibbs distribution, and the first term becomes its product with a function 
\begin{equation}
R_{\kappa\alpha}=1-p^{-\gamma}_{\kappa\alpha} 
\end{equation}
This function also appears under the integral sign, such that we can write the latter in an abbreviated version
\begin{equation}
\frac{S_\alpha\{p_{\kappa\alpha}(\beta)\}}{\kappa k_B}=P_{\kappa\alpha} R_{\kappa\alpha}-\int \frac{d\beta}{\beta}P_{\kappa\alpha} R_{\kappa\alpha}
\end{equation}
This is the relation between the probabilities of states $\alpha$ and their corresponding entropies. The sum over all $\alpha$ states gives the total entropy
\begin{equation}
\frac{S_{\kappa}(\beta)}{\kappa k_B}=1-\log(\beta U_0) -\big\langle p^{-\gamma}_{\kappa\alpha}(\beta)\big\rangle+\int \frac{d\beta}{\beta}\big\langle p^{-\gamma}_{\kappa\alpha}(\beta)\big\rangle +G(\{V\})
\end{equation}
in terms of the average probability, i.e. the expectation value of the probability raised to the power $-\gamma$. Again, the angular brackets indicate the probability weighted average over all states $\alpha$. $U_0$ is some normalizing thermal energy which to chose is arbitrary. The term containing it is of little importance.

This entropy is substantially more complicated than in ordinary classical statistical mechanics. Nevertheless it exhibits the relation between entropy and probability. It distinguishes the Olbert-Lorentzian entropy from Boltzmann-Gibbs-Shannon. 

\subsection{Boltzmann-Gibbs like form of the Olbert entropy}
Some insight can be obtained when considering the functional $R_{\kappa\alpha}$, writing it
\begin{equation}
R_{\kappa\alpha} = 1-\exp\big(-\gamma\log p_{\kappa\alpha}\big)
\end{equation}
Expanding the exponential yields to first order
\begin{equation}
R_{\kappa\alpha} = \gamma\log p_{\kappa\alpha}  +\mathrm{higher~order~terms}
\end{equation}
Except for the factor $\gamma$ this is just Boltzmann's micro-canonical entropy which, after multiplication with the probability and summation respectively integration yields the classical expression for the average entropy. {From this equivalence we conclude that in the Olbert entropy $S_{\alpha}\{p_{\kappa\alpha}(\beta)\}$ the functional $R_{\kappa\alpha}$ plays exactly the role of Boltzmann's micro-canonical entropy.} However, in Boltzmann theory the logarithm of the probability is just the inverse of the Boltzmann factor of the energy of state $\alpha$, with the energy $E_\alpha$ expressed in terms of the probability $p_\alpha$. This is also exactly the meaning of the functional $R_{\kappa\alpha}\{p_{\kappa\alpha}\}$, {which enables us to formulate the general}
\paragraph*{\textbf{\textit{{Theorem}}}}
\emph{{Let $p_\alpha(\beta, E_\alpha)=f_\alpha(\beta,E_\alpha)$ be the properly defined probability of a micro-canonical state $E_\alpha$, and $F\{p_{\alpha}\} = E_\alpha(p_\alpha)$ the inverse of $f_\alpha$. Then, up to some numerical factors, the micro-canonical entropy $S_\alpha$ of the state $\alpha$, expressed in terms of  the probability $p_\alpha$, is defined as $S_\alpha\{p_\alpha\}\propto F\{p_\alpha\}$; and the mean entropy $S\equiv\langle S_\alpha\rangle$ of the micro-canonical system, given by the sum over all probability-weighted states $\alpha$, is obtained in the form}}
\begin{equation}
{S\propto \sum_\alpha p_\alpha S_\alpha\{p_\alpha\} \propto \sum_\alpha p_\alpha F_\alpha\{p_\alpha\}}
\end{equation}
\emph{{if only the inverse functional $F\{p_\alpha\}$ exists and can be given either analytically or numerically. This formula is the general prescription of calculating the entropy in the micro-canonical state.}}

Let us, for convenience, discuss just the leading first order terms in the above expression for the Olbert entropy, assuming for our purposes of understanding that the higher order terms do not substantially contribute, which in general might not always be true. It then follows from Eq. (\ref{eq-entr}) that
\begin{equation}\label{eq-entr-1}
\frac{S_\alpha\{p_{\kappa\alpha(\beta)}\}}{\gamma\kappa k_B}=P_{\kappa\alpha}(\beta)\big(\log P_{\kappa\alpha}+\log Z_\kappa\big)- \int \frac{d\beta}{\beta} P_{\kappa\alpha}\big(\log P_{\kappa\alpha}+\log Z_\kappa\big) + \dots
\end{equation}
Except for the difficulty with the integral term, the first terms look about familiar. However, interestingly, this holds for the unsummed entropy. Summation then leads to the average entropy
\begin{equation}
\frac{S_\kappa}{|\gamma|\kappa k_B} =\langle\log P_\kappa(\beta)\rangle+\langle\log Z_\kappa\rangle-\int\frac{d\beta}{\beta}\big[\langle\log P_\kappa(\beta)\rangle+\langle\log Z_\kappa\rangle\big]
\end{equation}
It reproduces the logarithmic dependence on the mean logarithm of the partition function in the second term. The first term also reproduces the classical dependence on the logarithm of the mean probability $P_\kappa$. Further discussion is however less transparent, and the role of any higher order terms in the expansion as well as the structure of the integral term obscure its interpretation. In this form, however, we may conclude that the $\kappa$-generalization of classical statistical mechanics maintains its basic structure at least to lowest order. In any case it becomes clear that the {Olbert-Lorentzian generalization can be justified in its application to micro-canonical and, after proper extension to include the dependence on particle number, also to canonical systems. This is very satisfactory as it gives Olbert-Lorentzian statistical mechanics and the resulting Olbert-$\kappa$ distribution a physically justified place in the treatment of many-particle systems like high temperature plasmas. The different expressions for the entropies are then available for the proper description of the evolution of such states in thermal equilibrium as well as in non-equilibrium. }

\subsection{Quantum considerations} 
In this subsection we, for completeness, though just briefly, touch on the quantum extensions of Lorentzian entropies. We argued above that there is no zero temperature limit of the Lorentzian statistics.  This holds generally. Fermi statistics in addition inhibits correlations in the sense that any states $\alpha$ could be occupied by more than one particle. Hence, correlations involved in $\kappa$ can only be of the nature of entanglements. Hence, in addition to our finding that the state $T=0$ is principally excluded, this additional restriction categorically excludes application to Fermi systems other than entanglement of two particles of opposing spins. {Below we briefly consider this case.}

What concerns Bose statistics, the latter restriction is relaxed. States can obey arbitrary occupation numbers. Hence, high energy states can exist. Then, one will be able to find an appropriate expression for the Bose entropy {which we will provide in a follow-up communication as this requires another lengthy derivation which goes beyond the present note. }

We just briefly mention another interesting quantum case resulting in Fermi systems, the entanglement or von Neumann entropy \citep{vonneumann1955}. It is defined as
\begin{equation}
S_{vN}= \mathrm{Trace}\big(\rho\log\rho)
\end{equation}
where $\rho=\sum_\alpha|\psi_\alpha\rangle\langle\psi_\alpha|$ is the average scattering matrix in a quantum system, and Trace is its trace. Clearly if all $\psi_\alpha$ are true eigenstates of the entire system, $\rho=0$. Then the system is in its own eigenstate, and no entropy is produced. Otherwise, the entropy results from superposing all eigenstates $|\alpha\rangle$ of its components, yielding $\rho=\sum_\alpha\eta_\alpha|\alpha\rangle\langle\alpha|$ which contains all the irreversible interactions encoded in the superpositions of eigenstates of the components which contribute to the common wave function of the entire many particle system. Intuitively this is clear because all the different phases of the components will mix; the common wave function, being the superposition of all individual or grouped particle wave functions, will by no means become an eigenstate of the system. This is very frequently misunderstood when talking about fluid models of quantum theory and identifying the density with the expectation value of the wave function. In a quantum mechanical Olbert $\kappa$ system, where the particles are correlated and by some interaction mechanism are bunched together one may even expect that the scattering matrix contains non-diagonal terms  indicating dissipation. One such mechanism is entanglement between two prepared Fermions of opposite spin. By it two particles (electrons in the same state but of different spin) are bound together in their common behaviour. They are subject to von Neumann's entropy. If the entanglement can be encoded into a parameter $\kappa$, then its entropy may be conjectured to become
 \begin{equation}
S_{vN\kappa}\sim \mathrm{Trace} \big( \rho_\kappa\log\rho_\kappa^{1-\gamma}\big)=(1-\gamma)\mathrm{Trace} \big( \rho_\kappa\log\rho_\kappa\big)
\end{equation}
and one has $\rho_\kappa=\sum_\alpha|\psi_{\kappa\alpha}\rangle\langle\psi_{\kappa\alpha}|=\sum_\alpha\eta_{\kappa\alpha}|\alpha_\kappa\rangle\langle\alpha_\kappa|$. The $\kappa$ wave function might, however, not be known a priori. Since entanglement  applies to electrons, or in general Fermions, which by our above reasoning are not subject to Lorentzian statistics, then in $\kappa$-statistics it would apply to the bosonic property of paired electrons of opposite spin and must thus somehow, though not in an elucidated manner, relate to Boson-Lorentzian entropy of collectively grouped pairs like in superconductivity. If true, the parameter $\kappa$ appearing in the von Neumann entropy  then contains the physics of group entanglement. Otherwise $\kappa$ statistics does not apply in no manner to any entanglement, and no von Neumann-Lorentzian entropy exists.

\section{Conclusions}
In the present note we have undertaken the task of trying to understand what physically would be behind  Olbert-Lorentzian statistics. As the Olbert-$\kappa$ distribution function which belongs to it is well confirmed from a large number of observations mainly in space plasmas, {this effort is needed to give a clue on its foundations}. Applying statistical mechanical reasoning we have obtained expressions for the entropy as a functional of energy and also as functional of probability of states. What is most interesting in such an approach, is that the Olbert entropy  $S_\kappa$ has an equivalent form to that in ordinary non-equilibrium statistical mechanics. {The Olbert entropy, however, contains additional terms which can be calculated in an iterative perturbation theoretical way. It is for this reason super-additive (or super-extensive if wanted),} a property which it has in common with $q$-statistics though being rather different. We have elucidated the main difference here. This means that in $\kappa$-systems, i.e. for instance in high temperature plasmas exhibiting Olbert-distributions, the particles are correlatively grouped together to behave collectively, thereby providing the collective contribution to entropy. Such correlations are implicit to the index $\kappa$ and indicate strong nonlinear couplings which are provided by interaction potentials which are mediated not by collisions but by excitation of waves. {It is thus not surprising if $\kappa$-distributions are found in turbulent dilute high temperature plasmas like the solar wind \citep{goldstein2005}, near collisionless shock waves \citep{balogh2013}, Earth's bow shock \citep{eastwood2005}, the magnetosheath \citep{lucek2005}, at the boundaries of the heliosphere and astrospheres \citep{scherer2020},} where various types of waves can be excited as both, eigenmodes or sidebands, which even occupy the evanescent branches of the dielectric response function causing a continuous almost featureless power spectrum of fluctuations which is typical for well developed turbulence. One may, therefore, expect that the statistical mechanics underlying well developed collisionless turbulence will become kind of {Olbert-Lorentzian}  in terms of the probability distribution. The precise relation between these interactions and the particular value of the parameter $\kappa$ is still open to investigation. The consideration of entropy given here only shows its micro-canonical statistical-mechanical effect. 





\end{document}